\documentclass[prl,reprint,letterpaper,twocolumn,10pt,superscriptaddress,showpacs]{revtex4-1}

\usepackage{graphicx}
\usepackage{amssymb,amsfonts,amsmath,bm,mathtools}
\usepackage{tikz}

\newcommand{\commut}[2]{\left[{#1},{#2}\right]}

\newcommand{\ket}[1]{\left\lvert{#1}\right\rangle}
\newcommand{\norm}[1]{\left\lvert{#1}\right\rvert}
\newcommand{\mom}[1]{\left\langle{#1}\right\rangle}

\newcommand{\sgn}{\,\mathrm{sgn}}

\makeatletter
\renewcommand\thefigure{\@arabic\c@figure}
\renewcommand\fnum@figure{\figurename~\thefigure}
\newcommand\colorcapon{\renewcommand\fnum@figure{\figurename~\thefigure~(color online)}}
\newcommand\colorcapoff{\renewcommand\fnum@figure{\figurename~\thefigure}}
\makeatother

\begin{document}
\title{Unidimensional time domain quantum optics}
\author{St\'{e}phane Virally}
\email{Stephane.Virally@USherbrooke.ca}
\affiliation{Institut Quantique, D\'{e}partement de Physique, Universit\'{e} de Sherbrooke, Sherbrooke, Qu\'{e}bec J1K 2R1, Canada}
\affiliation{femtoQ Lab, D\'{e}partement de G\'{e}nie Physique, \'{E}cole Polytechnique de Montr\'{e}al, Montr\'{e}al, Qu\'{e}bec H3T 1JK, Canada}
\author{Bertrand Reulet}
\affiliation{Institut Quantique, D\'{e}partement de Physique, Universit\'{e} de Sherbrooke, Sherbrooke, Qu\'{e}bec J1K 2R1, Canada}
\date{\today}
\begin{abstract}
    Choosing the right first quantization basis in quantum optics is critical for the interpretation of experimental results. The usual frequency basis is, for instance, inappropriate for short, subcycle waveforms. Deriving first quantization in time domain shows that the electromagnetic field is not directly proportional, nor even causally related, to the photonic field (the amplitude probability of a photon detection). We derive the relation between the two and calculate the statistics of the electromagnetic field for specific states in time domain, such as the single photon Fock state. We introduce the dual of the Hamiltonian in time domain and extend the concept of quadratures to all first quantization bases.
\end{abstract}
\pacs{72.70.+m, 42.50.Ar}

\maketitle
\emph{Introduction.} 
Quantization of the free electromagnetic (EM) field is carried out in two steps: i)~first quantization~\cite{Bohr1913a} sorts various parts of the field into modes and associates a complex amplitude to each of them; ii)~second quantization~\cite{Dirac1927} upgrades the status of the amplitude and its complex conjugate to that of a couple of hermitian conjugate operators obeying bosonic commutation relations.

\begin{figure}
    \centering
    \includegraphics[width=0.48\textwidth]{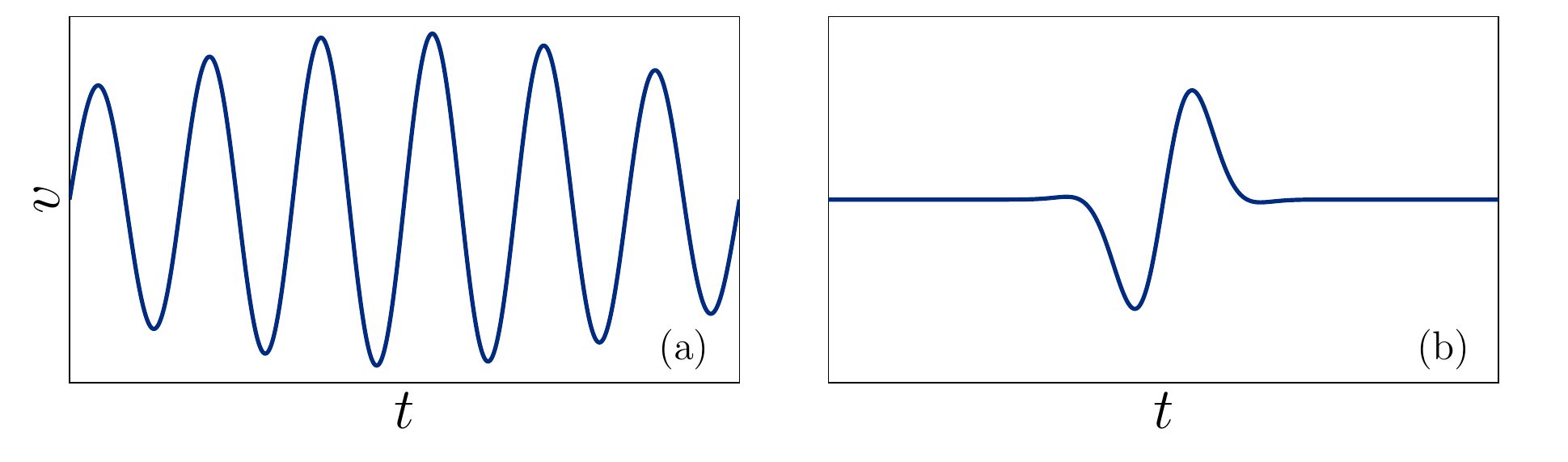}
    \caption{Voltage vs. time for quasi monochromatic (a) and quasi temporally localized (b) signals. First quantization in frequency is well-suited for the signal on the left. It is not the case for the ultrabroadband signal on the left, for which first quantization in time is more natural.}
    \label{f}
\end{figure}

Textbook treatments of first quantization enclose the free electromagnetic (EM) field in a fictitious cavity and expands it in terms of resonant frequencies~\cite{Scully1997,Loudon2000,Walls2008}. This is first quantization in the frequency domain, and it is perfectly appropriate for quasi-monochromatic modes of light, such as the modes of a laser. With shorter pulses and larger spectra, the same type of first quantization can be used, as long as there exists a well defined central carrier frequency (see Fig.~\ref{f}, a). However, one might wonder whether first quantization in the frequency domain is the right basis for pulse widths of the order of one optical cycle (as in Fig.~\ref{f}, b).

Progress in ultrafast optics~\cite{Reid2016} has seen generation of femtosecond pulses of subcycle durations~\cite{Krauss2009,Wirth2011}, thus motivating different approaches in the treatment of the quantum properties of such radiation. Recent measurements of mid-infrared fields have demonstrated experimental capability to subcycle sample quantum fields localized in space and time~\cite{Riek2015,Benea-Chelmus2016,Riek2017}. Subcycle signals can also readily be generated, propagated and measured in the microwave part of the electromagnetic spectrum. The only restriction with the exploration of such signals in the microwave quantum regime is that the temperature of the conductors must be lowered to a few tens of mK, so that blackbody radiation, i.e. thermal noise, does not overwhelm the signal ($k_\textrm{B}T/h\simeq$21GHz for $T$=1K). This regime can be easily achieved in dilution refrigerators, and quantum optical properties of microwave signals are the object of intense scrutiny~\cite{Beenakker2001,*Beenakker2004,Zakka-Bajjani2007,*Zakka-Bajjani2010,Menzel2010,Bozyigit2010,Lang2013,DiCandia2014,Forgues2015,Mendes2015b,Grimsmo2016,Virally2016,*Simoneau2016,Westig2017a,Goetz2017,Gu2017}, in particular within the framework of circuit quantum electrodynamics~\cite{DaSilva2010}.

In the narrow-bandwidth regime, there is a direct link between energy and photon number: the mean energy is almost $\mom{n}h\nu$, with $\nu$ the center frequency and $\mom{n}$ the average number of photons. As a result, the photon is sometimes described as the quantum of energy of the EM field. But in the narrow-time window regime, such a link does not exist, which can lead to the idea that photon numbers are ill defined. This contrasts with the viewpoint expressed in earlier seminal papers~\cite{Sipe1995,Smith2007}, where a photon is not linked to a specific energy. We take the same point of view and assert that a good operational definition of the photon is simply that of a ``click'' on a cascaded photodetector. Since this type of event can be observed, there must be a probability amplitude associated to it. We start with this amplitude and show that its link with the electromagnetic field is more complicated than what first quantization in frequency domain seems to indicate. The object of this Letter is to answer questions like: how many photons are expected in the short pulse shown in Fig.~\ref{f}(b)? What are their expected energies and times of arrival? What is the shape of the electric field or voltage associated with a single photon created at time $t$?

This Letter is organized as follows. We first describe first quantization for unidimensional waveguides in space and time domains. We start with operators for the amplitude of probability of observing a ``click'' on a photodetector (the ``photonic field''). We introduce a new observable, the weighted time of arrival, which is a time domain dual of the Hamiltonian. We also introduce the notion of photon packets. In a second part, we apply our theory to microwave signals and show in particular that the link between the photonic and electromagnetic fields is not causal. As a direct application, we predict the shape of the expected moments of the voltage associated with specific photonic states, including the single photon Fock state. We also extend the definition of quadratures to the time domain using the Hilbert transform. We finally show that the choice of first quantization basis must be tailored to the experimental setup to adequately explain results.

\vspace{.5\baselineskip}
\emph{First quantization in space.}
We consider a unidimensional waveguide parameterized by a position $z$. We are interested in a signal propagating the information that a ``click'' should be observed at position $z$ and time $t$. Classically, such a signal is a real function $s(z,t)$, which we assume propagate at velocity $v$, the speed of information in the waveguide. We also change the notation and define a time-like variable $\tau=z/v$ instead of the original position parameter. We thus use $s(\tau,t)$, with $\tau$ a time-like label for position and $t$ the real time.

Considering the signal in a quantum setting, we start with first quantization in $\tau$-time domain, and for the moment set $t=0$.
Although photons have been described as non localizable~\cite{Newton1949}, several arguments have been made that photonic modes, more broadly considered, can be localized in some settings~\cite{Fronsdal1959,Amrein1969,Acharya1960a}, and in particular in the unidimensional case (coaxial cable, waveguide, optical fiber) that we are exploring.
We thus assume that there exists a complex wave function $a(\tau)$ with the property that $\norm{a(\tau)}^2$ gives the probability of detecting a photon (a click on a perfect detector) at location $\tau$ and time $t=0$. Second quantization promotes that wave function and its complex conjugate to the status of hermitian conjugate ladder operators $\bm{a}_\tau$ and $\bm{a}^\dagger_\tau$. It is appropriate to treat $\tau$ as a continuous mode index rather than a variable, and this is reflected in the notation.

We then go to the spatial frequency domain, and define the Fourier transform $\bm{a}_\nu$ of $\bm{a}_\tau$, and its hermitian conjugate $\bm{a}^\dagger_\nu$, as
\begin{equation}
    \label{aleftright}
    \bm{a}_\nu=\int_{-\infty}^{+\infty}d\tau\;\bm{a}_\tau\,e^{-i2\pi\nu\tau};\,\bm{a}^\dagger_\nu=\int_{-\infty}^{+\infty}d\tau\;\bm{a}^\dagger_\tau\,e^{i2\pi\nu\tau},
\end{equation}
where $\nu$ is a spatial frequency, the equivalent of a wavevector. Positive (negative) spatial frequencies represent signals propagating in the $+z$ (-$z$)direction. Note that $\bm{a}^\dagger_\nu$ is not the Fourier transform of $\bm{a}^\dagger_\tau$.

We impose the usual bosonic commutation relation
\begin{equation}
    \label{commutnu}
    \commut{\bm{a}_\nu}{\bm{a}^\dagger_{\nu'}}=\delta(\nu-\nu')\Leftrightarrow
    \commut{\bm{a}_\tau}{\bm{a}^\dagger_{\tau'}}=\delta(\tau-\tau').
\end{equation}
The $\bm{a}_\tau$ modes are localized and thus are not the usual (propagating) photonic modes of quantum optics. They are in fact a superposition of $\pm\tau$-propagating directional modes $\bm{a}_{\tau,\pm}$
\begin{equation}
    \label{apm}
    \begin{split}
        \bm{a}_\tau&=\int_{-\infty}^{0}d\nu\;\bm{a}_\nu\,e^{i2\pi\nu\tau}+\int_{0}^{+\infty}d\nu\;\bm{a}_\nu\,e^{i2\pi\nu\tau};\\
        &\equiv\bm{a}_{\tau,-}+\bm{a}_{\tau,+}=\sum_\sigma\bm{a}_{\tau,\sigma},
    \end{split}
\end{equation}
where, $\sigma$ stands for ``$+$'' or ``$-$" (alternatively, $\pm1$ where required). The usual propagating modes~\cite{Yurke1984,Devoret2016} are the Fourier transforms of the $\bm{a}_{\tau,\pm}$, which could be referred to as $\bm{a}_{\nu,\pm}$, but Eq.~\eqref{apm} clearly shows that they simply are the $\bm{a}_\nu$ for $\nu$ positive or negative. This stems from the fact that $\nu$ is really a wavevector. 

\vspace{.5\baselineskip}
\emph{Time evolution and first quantization in time.}
Under the free-photon Hamiltonian
\begin{equation}
    \mathbf{H}=\int_{-\infty}^{+\infty}d\nu\;\frac{h\!\norm{\nu}}{2}\left[\bm{a}_{\nu}^\dagger\bm{a}_{\nu}+\bm{a}_{\nu}\bm{a}_{\nu}^\dagger\right],
\end{equation}
the ladder operators evolve in time as
\begin{equation}
    \label{anusigmat}
    \begin{split}
        &\bm{a}_{\nu}\rightarrow\bm{a}_{\nu}(t)=\bm{a}_{\nu}\,e^{-i2\pi\norm{\nu}t};\\
        &\bm{a}_{\nu}^\dagger\rightarrow\bm{a}_{\nu}^\dagger(t)=\bm{a}_{\nu}^\dagger\,e^{i2\pi\norm{\nu}t},
    \end{split}
\end{equation}
from which the operators $\bm{a}_{\tau,\sigma}(t)$ are straightforwardly derived. We note that contrary to the usual treatment of quantum circuit theory~\cite{Yurke1984}, we do not identify $t$, the real time, and $\tau$, the spatial index~\cite{Devoret2016}. It is indeed perfectly possible to observe, for instance, quantities varying in time at a single position.

As directional modes cannot be fully localized, the $\bm{a}_{\tau,\sigma}(t)$ operators do not obey bosonic commutation relations of the form~\eqref{commutnu}. However, they obey the physically meaningful relations~\footnotemark[1]
\begin{equation}
    \label{commuthalftau}
    \commut{\bm{a}_{\tau,\sigma}(t)}{\bm{a}_{\tau',\sigma'}^\dagger(t')}=\delta_{\sigma,\sigma'}\,\delta[(t-t')-\sigma(\tau-\tau')].
\end{equation}

This striking result links two space-time events (the creation of a photon at time $t$ and position $\tau$ and an annihilation at time $t'$ and position $\tau'$) on the condition that they are light-like separated, i.e. that they obey causality. 

The Fourier transform of $\bm{a}_{\tau,\sigma}(t)$, with respect to real time, is $\bm{a}_{\tau,\sigma}(f)$. Its hermitian conjugate is $\bm{a}_{\tau,\sigma}^\dagger(f)$. Using the relations~\eqref{apm}, we find that $\bm{a}_{\tau,\sigma}^{(\dagger)}(f)=\mathbf{0}$ when $f<0$. Thus, the real frequency of propagating modes is naturally always positive.

The operators $\bm{a}_{\tau,\sigma}^{(\dagger)}(t)$ and $\bm{a}_{\tau,\sigma}^{(\dagger)}(f)$ are the most relevant for the description of experiments, as they relate to detection at a fixed location, in real time and frequency domains. We focus the remainder of the text on these operators. Fixing $\tau$ and $\sigma$, we simply write $\bm{a}^{(\dagger)}\equiv\bm{a}_{\tau,\sigma}^{(\dagger)}$. In this Hilbert subspace of photons traveling in a single direction and measured at a single location, we get the usual Hamiltonian of quantum optics written in frequency domain,
\begin{equation}
    \label{hbof}
    \mathbf{H}_{\tau,\sigma}\equiv\bm{H}=\int_0^{+\infty}df\;\frac{hf}{2}\left[\bm{a}^\dagger(f)\bm{a}(f)+\bm{a}(f)\bm{a}^\dagger(f)\right],
\end{equation}
and, having put time and frequency domains on the same footing, we are lead to introduce the dual ``weighted time'' operator, in time domain,
\begin{equation}
    \label{zetheta}
    \bm\theta=\int_{-\infty}^{+\infty}dt\;\frac{t}{2}\left[\bm{a}^\dagger(t)\,\bm{a}(t)+\bm{a}(t)\bm{a}^\dagger(t)\right],
\end{equation}
which is written simply in time domain and is the observable associated with the weighted time of arrival of photons (measured by an infinitely fast detector). The eigenvectors of the Hamiltonian are states for which the energy remains the same when time varies. Eigenvectors of $\bm{\theta}$ should then be states with identical time characteristics (potentially in a general way, including periodicity), independent of their energy (like the periods of an harmonic oscillator).

In the same way that measuring the energy yields no information on the time of arrival (as a perfect measurement requires an infinite amount of time), measuring the time of arrival of photons yields no information on their energy (as a perfect measurement requires an infinite amount of bandwidth). This is reflected in the duality of the $\bm{H}$ and $\bm{\theta}$ observables, as their commutator is~\footnotemark[1]
\begin{equation}
    \label{zecommut}
    \commut{\bm{H}}{\bm{\theta}}=i\hbar\bm{N},
\end{equation}
where $\bm{N}$ is the total photon number operator.
The commutator~\eqref{zecommut} leads to the uncertainty relation
\begin{equation}
    \label{uncertainty}
    \sqrt{\mom{\Delta\bm{\theta}^2}\mom{\Delta\bm{H}^2}}\ge\frac{\hbar}{2}\mom{\bm{N}},
\end{equation}
where $\mom{\Delta\bm\Omega^2}$ is the variance and $\mom{\bm\Omega}$ the expected value of any observable $\bm\Omega$. Note that the commutator~\eqref{zecommut} is not $\commut{\bm{\theta}}{\bm{H}}=i\hbar\mathbb{I}$. Pauli showed that if such a relation existed, the energy spectrum could not be lower-bounded~\cite[p.~63]{Pauli1980}. But Pauli's argument does \emph{not} apply here.

In the ``particle/wave'' photon duality context, the usual wave packets stand on the wave (frequency) side. On the dual particle (time) side, we introduce photon packets with ladder operators
\begin{equation}
    \bm{A}=\int_{-\infty}^{+\infty}dt\;\phi^*(t)\,\bm{a}(t),
\end{equation}
where $\phi$ is an integral-normalized complex function that can be as narrow in time as is required to reflect the speed of a detection or emission process. These operators verify the usual bosonic commutator $\commut{\bm{A}}{\bm{A}^\dagger}=\mathbb{I}$ and can be used to generate photonic states (Fock, coherent, squeezed, \dots) in the usual manner~\cite{Scully1997,Loudon2000,Walls2008}.

\colorcapon
\begin{figure}[h]
    \centering
    \includegraphics[width=0.48\textwidth]{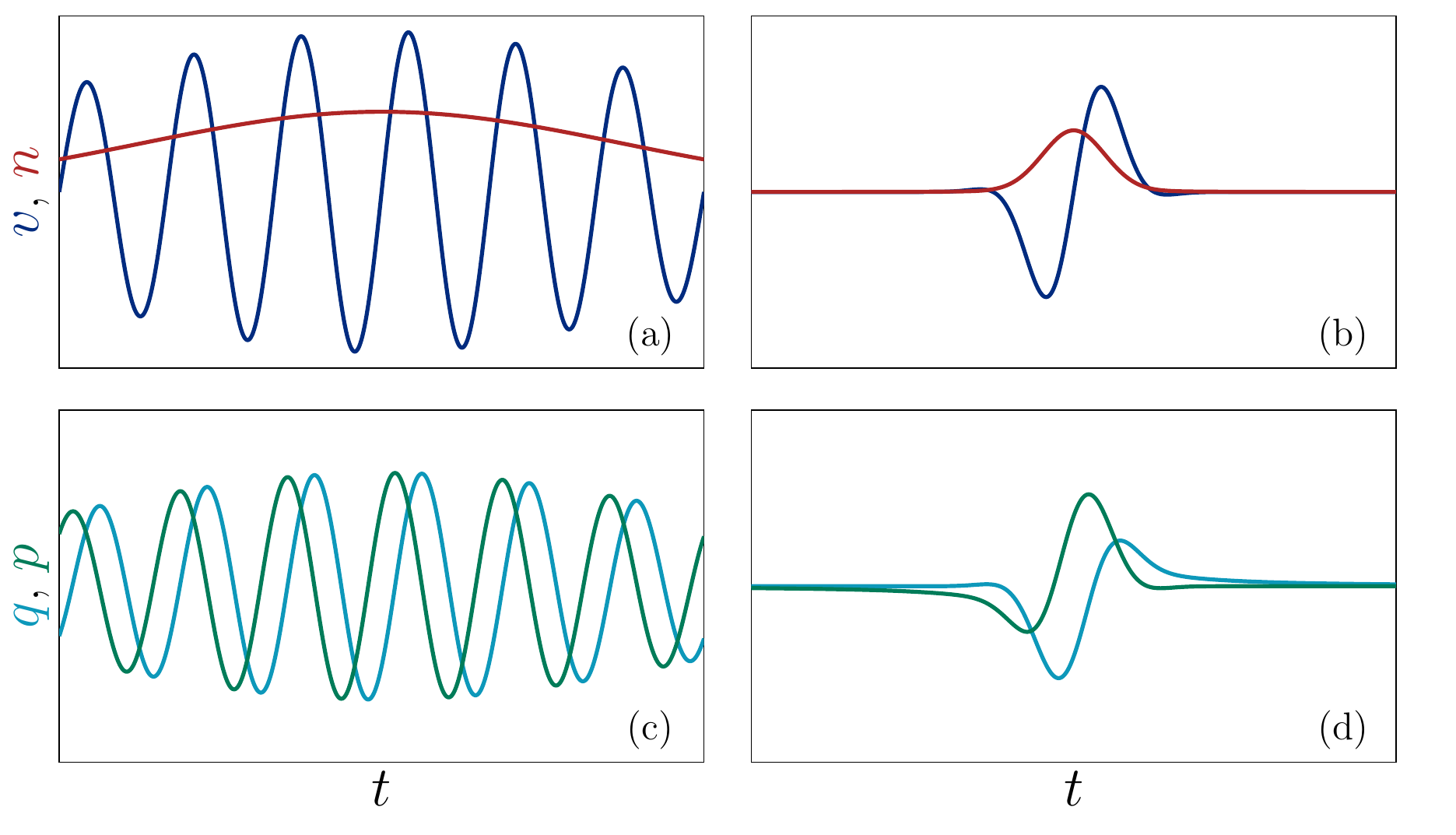}
    \caption{Temporal evolutions [arb. units] of: (a,b) voltage signals (blue), instantaneous photon fluxes (red), (c,d) $p$ quadratures (green) and $q$ quadratures (light blue) in the two regimes considered in Figure~\ref{f}.}
    \label{f2}
\end{figure}
\colorcapoff

\vspace{.5\baselineskip}
\emph{Application to microwaves.}
A clear distinction must be made between the photon packets and the usual wave packets of quantum optics in frequency domain. The former are linked to photon emission/detection probability amplitudes, and the latter to energy emission/detection amplitudes. Photon packets are \emph{not} directly proportional to electromagnetic field components. In fact, we now show that they are not even \emph{causally related} to them.
We illustrate this fact in the microwave domain, where the main electromagnetic observable is the voltage~\cite{Yurke1984}
\begin{equation}
    \bm{v}(t)=-i\int_0^{+\infty}df\;\sqrt{\frac{Zhf}{2}}\left[\bm{a}(f)\,e^{-i2\pi ft}-\mathrm{h.c.}\right],
\end{equation}
with $Z$ the characteristic impedance of the transmission line, here taken as independent of frequency.
Expressed as a function of the $\bm{a}^{(\dagger)}(t)$ instead, the voltage operator is
\begin{equation}
    \bm{v}(t)=\int_{-\infty}^{+\infty}dt'\;\frac{-\sqrt{Zh}}{8\pi\norm{t'}^{3/2}}\left[\sgn(t')\bm{q}(t-t')+\bm{p}(t-t')\right],
\end{equation}
where $\bm{q}(t)$ and $\bm{p}(t)$ are the usual quadratures~\cite{Scully1997, Loudon2000, Walls2008} extended to time domain~\footnotemark[1] and $\sgn$ is the signum function.

Inversely, we have
\begin{align}
    \label{p_t}
    &\bm{p}(t)=\int_{-\infty}^{+\infty}dt'\;\sqrt{\frac{2}{Zh\norm{t'}}}\;\bm{v}(t-t');\\
    \label{q_t}
    &\bm{q}(t)=\int_{-\infty}^{+\infty}dt'\;\sqrt{\frac{2}{Zh\norm{t'}}}\;\sgn(t')\bm{v}(t-t').
\end{align}

Extended quadratures are Hilbert transforms of one another~\footnotemark[1]. This is remarkable but not entirely surprising, as the Hilbert transform is the operation that exchanges sine and cosine in the frequency domain. Quadratures of the signals of Fig.~\ref{f} are shown in the second row of Fig.~\ref{f2}.

Transforms~\eqref{p_t} and~\eqref{q_t} are non-local in time and necessitate both past and future amplitudes. There is thus no direct causal transformation from the electromagnetic field to the photonic field, and vice-versa. This is due to the non-local nature of the directional modes (even in time~\cite{Kim2000a}) and does not imply that relativistic causality is violated~\cite{Wheeler1945,*Wheeler1949,Cramer1986}. It means, for instance, that the instantaneous photon flux
\begin{equation}
    \label{n_t}
    \bm{n}(t)=\frac{1}{2}\left[\bm{q}^2(t)+\bm{p}^2(t)\right],
\end{equation}
cannot be measured on the fly by an oscilloscope and must be recovered using past and future input. Inversely, (squared) voltage and field values cannot be inferred from past photocounts alone. This is actually very much in line with the idea that quantum theories must be time symmetric~\cite{Schottky1921} and with the two-vector state formalism~\cite{Watanabe1955,Aharonov1964}, where the full quantum state is described at all times by a forward-propagating ket that is fully predicted by all past measurements, and a backward-propagating bra that is fully retrodicted by all future measurements.

\begin{figure}
    \centering
    \includegraphics[width=0.48\textwidth]{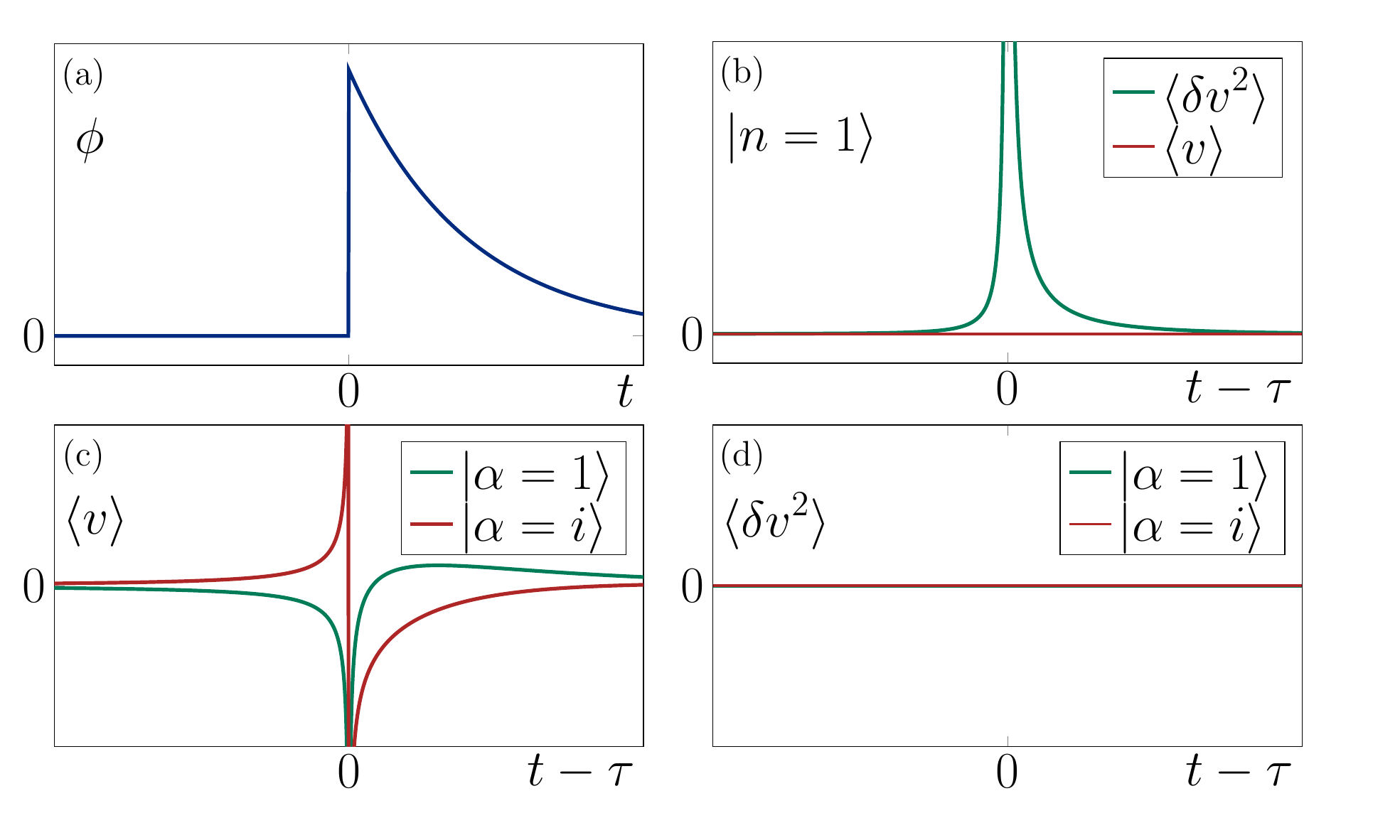}
    \caption{Illustrations of an exponentially decaying photon packet emitted at location $\tau=0$ and time $t=0$, with a characteristic time $\sigma_t$. Top: Probability amplitude~(a) $\phi$ (taken as real) for the photon packet, as a function of time $t$; voltage average and variance~(b), at location $\tau$, as a function of the time difference $t-\tau$, for the Fock state $\ket{n=1}$ ($\mom{v(t)}$ is zero at all times for all Fock states); the variance behaves as $Z\hbar/(4\pi\norm{t-\tau}\sigma_t)$ close to $t-\tau=0$; Bottom: Voltage average~(c) and variance~(d), as a function of the time difference $t-\tau$, for the coherent states $\ket{\alpha=1}$ and $\ket{\alpha=i}$. As expected, the fluctuations for any coherent state are equal to the sole vacuum fluctuations, which have been subtracted from all results. Time scales are identical on all figures.}
    \label{f3}
\end{figure}

However, the theory does predict statistical moments of the voltage or field values for well-defined photon packets. As an example, we calculate voltage moments for a single photon Fock state and for two quadrature-related coherent states with mean photon number equal to one. Results depend on the assumptions about the initial shape of the photon packet. Fig~\ref{f3} shows an example for a typical decay photon packet (this could apply for instance to a single emission event from a system excited at $t=0$). Additional cases are explored in the supplementary material~\footnotemark[1].

New information can also be gathered from voltage traces. For instance, Eq.~\eqref{zetheta} leads to the statistics of time of arrival. To better understand the physical meaning of $\bm\theta$, we consider a single pulse for which we can measure both $\bm\theta$ and $\bm{N}$, as they commute. Repeating the experiment, the quantity $\mom{\bm\theta}/\mom{\bm{N}}$ is the mean time of arrival of pulses, while $\sqrt{\mom{\Delta\bm\theta^2}}/\mom{\bm{N}}$ (which involves the two-time correlator $\mom{\bm{n}(t_1)\,\bm{n}(t_2)}$) is the uncertainty in the time of arrival, or jitter. In contrast, the pulse width (or intra-pulse time uncertainty) involves the quantity $\mom{\int dt\; t^2\,\bm{n}(t)}$.

Using definition~\eqref{n_t}, we find that the instantaneous photon flux associated with an infinite sinusoidal voltage is constant at all times, as expected. It is reasonable to expect, as shown in Fig.~\ref{f2} (a),(b), that the instantaneous photon flux \emph{resembles} the envelope of the electromagnetic pulse. However, since there is no causal link between the electromagnetic and photonic fields, it is absolutely possible for the photon flux to be at a maximum when the voltage is zero, as shown in Fig.~\ref{f2} (b).

\vspace{.5\baselineskip}
\emph{Conclusion.}
Both frequency domain and time domain first quantization limits are useful to understand the electromagnetic modes of a unidimensional waveguide. Photocount observable lie in-between the time and frequency domain and are not more naturally described in one limit over the other one. Measurement apparatuses should always dictate the best point of view. For instance, ultrabroadband quantum experiments with mesoscopic devices such as tunnel and Josephson junctions~\cite{Gabelli2008, Forgues2015, Simoneau2016, Westig2017a} are likely to be best modeled in time domain, where furthermore the ``weighted time'' operator (equivalent of the Hamiltonian in the frequency domain) yields statistics of photon time of arrival.

The photonic and electromagnetic fields are only related in a non-causal way. Hence, the probability to observe a ``click'' on a detector at a given time depends on non-local features of the electromagnetic field and can be maximal when the instantaneous voltage vanishes. The theory herein allows the prediction of all moments of the electromagnetic field from the initial first-quantization shape of the photon packet, and its second-quantized statistical moments. It also allows the reverse, i.e. the full counting statistics of photons given voltage of electric field moments. We expect that it will for instance prove useful in calculating outcomes of interactions between the electromagnetic field and the electron field in a conductor, especially over short times. It also  applies to 1d interactions with dielectrics, and could be extended to higher dimensions.

\vspace{\baselineskip}
\emph{Acknowledgements.}
The authors acknowledge fruitful discussions with Wolfgang Belzig, Guido Burkard, Michel Devoret, Nicolas Godbout, Udson Mendes and Denis Seletskiy. This work was supported by the Canada Excellence Research Chairs program, Canada NSERC, Qu\'{e}bec MEIE, Qu\'{e}bec FRQNT via INTRIQ, Universit\'{e} de Sherbrooke via EPIQ, and Canada Foundation for Innovation.

\bibliographystyle{apsrev4-1}
\bibliography{UDTDQOv2_main}

\footnotetext[1]{See supplementary material.}

\end{document}


\title{Supplementary material for ``Unidimensional Time Domain Quantum Optics''}
\author{St\'{e}phane Virally}
\email{Stephane.Virally@USherbrooke.ca}
\affiliation{Institut Quantique, D\'{e}partement de Physique, Universit\'{e} de Sherbrooke, Sherbrooke, Qu\'{e}bec J1K 2R1, Canada}
\affiliation{femtoQ, D\'{e}partement de G\'{e}nie Physique, Polytechnique Montr\'{e}al, Montr\'{e}al, Qu\'{e}bec H3T 1JK, Canada}
\author{Bertrand Reulet}
\affiliation{Institut Quantique, D\'{e}partement de Physique, Universit\'{e} de Sherbrooke, Sherbrooke, Qu\'{e}bec J1K 2R1, Canada}
\date{\today}
\pacs{72.70.+m, 42.50.Ar}
\maketitle

\section{Definitions}
Throughout, we use $\int\equiv\int_{-\infty}^{+\infty}$, $\int_+=\int_0^{+\infty}$,$\int_-=\int_{-\infty}^0$.\\

We consider a unidimensional waveguide parameterized by a position $z$, which we replace, for practical purposes, by a time-like variable $\tau=z/v$, where $v$ is the speed of light (and information) in the waveguide. We also define $\sigma$ as either $+$ or $-$, and, where warranted, as either $+1$ or $-1$. The dual of $\tau$ is the spatial frequency $\nu$, akin to a wavevector.

\section{Ladder operators}
\subsection{All ladder operators of a unidimensional waveguide}
The unidimensional waveguide can be populated by bosonic traveling modes. Second quantization of these modes, in the continuum limit of first quantization, yields the ladder operators
\begin{equation}
    \bm{a}_\nu,\bm{a}_\nu^\dagger.
\end{equation}


Ladder operators in the $\tau$-time domain are the \emph{localized} modes
\begin{equation}
    \label{atau}
    \begin{matrix*}[l]
        &\bm{a}_{\tau}=\bm{a}_{\tau,+}+\bm{a}_{\tau,-};&
        &\bm{a}_{\tau}^\dagger=\bm{a}_{\tau,+}^\dagger+\bm{a}_{\tau,-}^\dagger,
    \end{matrix*}
\end{equation}
sums of the $\pm\tau$-\emph{propagating} modes
\begin{equation}
    \label{atausig}
    \begin{matrix*}[l]
        &\bm{a}_{\tau,\sigma}=\int_\sigma d\nu\;\bm{a}_\nu\,e^{i2\pi\nu\tau};&
        &\bm{a}_{\tau,\sigma}^\dagger=\int_\sigma d\nu\;\bm{a}_\nu^\dagger\,e^{-i2\pi\nu\tau}.
    \end{matrix*}
\end{equation}

The reciprocal relations are
\begin{equation}
    \label{anusig}
    \begin{matrix*}[l]
        &\bm{a}_{\nu,}=\int d\tau\;\bm{a}_{\tau}\,e^{-i2\pi\nu\tau};&
        &\bm{a}_{\nu}^\dagger=\int d\tau\;\bm{a}_{\tau}^\dagger\,e^{i2\pi\nu\tau}.
    \end{matrix*}
\end{equation}

\subsection{Hamiltonian and evolution in time}
The Hamiltonian of the electromagnetic field in the conductor is
\begin{equation}
    \bm{H}=\int d\nu\;\frac{h\!\norm{\nu}}{2}\left[\bm{a}_\nu^\dagger\bm{a}_\nu+\bm{a}_\nu\bm{a}_\nu^\dagger\right],
\end{equation}
where $h$ is Planck's constant.

Under the Hamiltonian, the operators evolve in time as
\begin{equation}
    \label{timeevolve}
    \begin{matrix*}[l]
        &\bm{a}_\nu(t)=\bm{a}_\nu\,e^{-i2\pi\norm{\nu}t};&
        &\bm{a}_\nu^\dagger(t)=\bm{a}_\nu^\dagger\,e^{i2\pi\norm{\nu}t};\\
        &\bm{a}_{\tau,\sigma}(t)=\bm{a}_{\tau,\sigma}\,e^{-i2\sigma\pi\nu t};&
        &\bm{a}_{\tau,\sigma}^\dagger(t)=\bm{a}_{\tau,\sigma}^\dagger\,e^{i2\sigma\pi\nu t}.\\
    \end{matrix*}
\end{equation}

In (time-related) frequency domain, we have
\begin{equation}
    \label{af}
    \begin{matrix*}[l]
        &\bm{a}_{\nu}(f)=\int dt\;\bm{a}_{\nu}(t)\,e^{i2\pi ft};&
        &\bm{a}_{\nu}^\dagger(f)=\int dt\;\bm{a}_{\nu}^\dagger(t)\,e^{-i2\pi ft};\\
        &\bm{a}_{\tau,\sigma}(f)=\int dt\;\bm{a}_{\tau,\sigma}(t)\,e^{i2\pi ft};&
        &\bm{a}_{\tau,\sigma}^\dagger(f)=\int dt\;\bm{a}_{\tau,\sigma}^\dagger(t)\,e^{-i2\pi ft}.
    \end{matrix*}
\end{equation}

Using Eqs.~\eqref{atau},~\eqref{atausig},~\eqref{anusig} on these definitions yields
\begin{equation}
    \bm{a}_{\nu}(f)=\bm{a}_{\nu}^\dagger(f)=\bm{a}_{\tau,\sigma}(f)=\bm{a}_{\tau,\sigma}^\dagger(f)=\bm0\quad\forall f<0,
\end{equation}
which means that all modes naturally only admit only positive (time-related) frequencies. Thus, the reciprocal transforms are
\begin{equation}
    \label{at}
    \begin{matrix*}[l]
        &\bm{a}_{\nu}(t)=\int_+ df\;\bm{a}_{\nu}(f)\,e^{-i2\pi ft};&
        &\bm{a}_{\nu}^\dagger(t)=\int_+ df\;\bm{a}_{\nu}^\dagger(f)\,e^{i2\pi ft};\\
        &\bm{a}_{\tau,\sigma}(t)=\int_+ df\;\bm{a}_{\tau,\sigma}(f)\,e^{-i2\pi ft};&
        &\bm{a}_{\tau,\sigma}^\dagger(t)=\int_+ df\;\bm{a}_{\tau,\sigma}^\dagger(f)\,e^{i2\pi ft}.
    \end{matrix*}
\end{equation}

\subsection{Commutators}
The canonical commutator
\begin{equation}
    \label{commutanu}
    \commut{\bm{a}_\nu}{\bm{a}^\dagger_{\nu'}}=\delta(\nu-\nu')
\end{equation}
yields
\begin{equation}
    \commut{\bm{a}_\tau}{\bm{a}^\dagger_{\tau'}}=\delta(\tau-\tau').
\end{equation}

Eqs.~\eqref{timeevolve} yield
\begin{equation}
    \label{commutanusigt}
    \commut{\bm{a}_{\nu}(t)}{\bm{a}_{\nu'}^\dagger(t')}=\delta(\nu-\nu')\,e^{-i2\pi\sigma\norm{\nu}(t-t')},
\end{equation}
and
\begin{equation}
    \label{commutatausigt}
    \commut{\bm{a}_{\tau,\sigma}(t)}{\bm{a}_{\tau',\sigma'}^\dagger(t')}=\delta_{\sigma,\sigma'}\,\delta[(t-t')-\sigma(\tau-\tau')].
\end{equation}
The meaning of this commutator is relatively clear, as it invokes the finite speed of light in the waveguide and links creation and annihilation events in spatial modes located on the same light-like world line.

For the localized modes, we get
\begin{equation}
    \label{commutataut}
    \commut{\bm{a}_\tau(t)}{\bm{a}_{\tau'}^\dagger(t')}=\delta[(t-t')-(\tau-\tau')]+\delta[(t-t')+(\tau-\tau')].
\end{equation}

Finally, using Eqs.~\eqref{atau},~\eqref{af}, we find
\begin{equation}
\label{commutatausigf}
    \commut{\bm{a}_{\tau,\sigma}(f)}{\bm{a}_{\tau',\sigma'}^\dagger(f')}=\delta_{\sigma,\sigma'}\,\delta(f-f')\,\,e^{i\sigma2\pi f(\tau-\tau')};
\end{equation}
and
\begin{equation}
    \label{commutatauf}
    \commut{\bm{a}_\tau(f)}{\bm{a}_{\tau'}^\dagger(f')}=2\;\delta(f-f')\cos[2\pi f(\tau-\tau')].
\end{equation}

\section{Modes observed at a single location}
In the following, we select a specific $(\tau,\sigma)$. Thus, we observe modes at a single location. We use the shorthand
\begin{equation}
    \begin{matrix*}[l]
        &\bm{a}\equiv\bm{a}_{\tau,\sigma};&
        &\bm{a}^\dagger\equiv\bm{a}_{\tau,\sigma}^\dagger.
    \end{matrix*}
\end{equation}

All observations are functions of $t$ and/or $f$.

The Hamiltonian of this Hilbert subspace is
\begin{equation}
    \bm{H}=\int_+ df\;\frac{hf}{2}\left[\bm{a}^\dagger(f)\bm{a}(f)+\bm{a}(f)\bm{a}^\dagger(f)\right].
\end{equation}

By way of similarity, we define the weighted average time operator as
\begin{equation}
    \bm\theta=\int dt\;\frac{t}{2}\left[\bm{a}^\dagger(t)\bm{a}(t)+\bm{a}(t)\bm{a}^\dagger(t)\right].
\end{equation}

Making use of the commutator~\eqref{commutatausigf}, we have
\begin{equation}
    \begin{split}
        &\phantom{=}\commut{\bm\theta}{\bm{H}}\\
        &=\frac{h}{4}\int_+ df\,df_1\,df_2\int dt\;\commut{\bm{a}^\dagger(f_1)\bm{a}(f_2)+\bm{a}(f_2)\bm{a}^\dagger(f_1)}{\bm{a}^\dagger(f)\bm{a}(f)+\bm{a}(f)\bm{a}^\dagger(f)}f\,t\,e^{i2\pi(f_1-f_2)t}\\
        &=\frac{-i\hbar}{4}\int_+ df\,df_1\,df_2\;\commut{\bm{a}^\dagger(f_1)\bm{a}(f_2)+\bm{a}(f_2)\bm{a}^\dagger(f_1)}{\bm{a}^\dagger(f)\bm{a}(f)+\bm{a}(f)\bm{a}^\dagger(f)}f\,\delta'(f_1-f_2)\\
        &=\frac{i\hbar}{2}\int_+df_1\,df_2\left[\bm{a}^\dagger(f_1)\bm{a}(f_2)+\bm{a}(f_2)\bm{a}^\dagger(f_1)\right](f_1-f_2)\delta'(f_1-f_2)\\
        &=-i\hbar\int_+df\;\frac{1}{2}\left[\bm{a}^\dagger(f)\bm{a}(f)+\bm{a}(f)\bm{a}^\dagger(f)\right]\\
        &=-i\hbar\bm{N}
    \end{split}
\end{equation}
with $\bm{N}$ the total photon number operator for the modes (including vacuum contributions).

The associated uncertainty relation is
\begin{equation}
    \sqrt{\mom{\Delta\bm\theta^2}\mom{\Delta\bm{H}^2}}\ge\frac{\hbar}{2}\mom{\bm{N}}.
\end{equation}

\subsection{Voltage operator}
The voltage operator is
\begin{equation}
    \begin{split}
        \bm{v}(t)&=i\sqrt{\frac{Zh}{2}}\int_+df\;\sqrt{f}\left[\bm{a}^\dagger(f)e^{i2\pi ft}-\bm{a}(f)e^{-i2\pi ft}\right]\\
        &=\frac{-i}{8\pi}\sqrt{\frac{Zh}{2}}\int dt'\;\frac{1}{\norm{t-t'}^{3/2}}\left\{\bm{a}^\dagger(t')[1+i\,\sgn(t-t')]-\bm{a}(t')[1-i\,\sgn(t-t')]\right\}
    \end{split}
\end{equation}

\subsection{Quadratures}
We extend the usual definition of quadratures to time domain,
\begin{equation}
    \bm{q}_\theta(t)\equiv\frac{\bm{a}^\dagger(t)e^{i\theta}+\bm{a}(t)e^{-i\theta}}{\sqrt{2}}.
\end{equation}

We also define the transforms
\begin{equation}
    \begin{split}
        &\bm{s}_+(t)=\int_+dt'\;\frac{\bm{v}(t-t')}{\sqrt{\norm{t'}}}=\int dt'\;\frac{\mathcal{H}(t')}{\sqrt{t'}}\;\bm{v}(t-t')\\
        &\bm{s}_-(t)=\int_-dt'\;\frac{\bm{v}(t-t')}{\sqrt{\norm{t'}}}=\int dt'\;\frac{\mathcal{H}(t')}{\sqrt{t'}}\;\bm{v}(t+t').
    \end{split}
\end{equation}

Using the Fourier transform
\begin{equation}
    \frac{\mathcal{H}(t)}{\sqrt{t}}=\int df\;\frac{1+i\,\sgn(f)}{2\sqrt{\norm{f}}}\,e^{-i2\pi ft},
\end{equation}
where $\sgn$ is the signum function, we get
\begin{equation}
    \begin{split}
        &\bm{s}_+(t)=\sqrt{\frac{Zh}{2}}\;\bm{q}_{\frac{\pi}{4}}(t);\\
        &\bm{s}_-(t)=\sqrt{\frac{Zh}{2}}\;\bm{q}_{\frac{3\pi}{4}}(t),
    \end{split}
\end{equation}
and
\begin{equation}
    \begin{split}
        &\frac{\bm{s}_+(t)+\bm{s}_-(t)}{2}=\int dt'\;\frac{\bm{v}(t-t')}{\sqrt{\norm{t'}}}=\sqrt{\frac{Zh}{2}}\;\bm{q}_{\frac{\pi}{2}}(t)\equiv\bm{p}(t);\\
        &\frac{\bm{s}_+(t)-\bm{s}_-(t)}{2}=\int dt'\;\frac{\sgn(t')\bm{v}(t-t')}{\sqrt{\norm{t'}}}=\sqrt{\frac{Zh}{2}}\;\bm{q}_{0}(t)\equiv\bm{q}(t).
    \end{split}
\end{equation}

Using the definition of the Hilbert transform
\begin{equation}
    f_h(t)=\mathrm{p.v.}\int dt'\;\frac{f(t-t')}{t'},
\end{equation}
where $\mathrm{p.v.}$ denotes the principal value, and using the Fourier transforms
\begin{equation}
    \begin{split}
        &\frac{1}{\pi t}=\int df\;i\,\sgn(f)\,e^{-i2\pi ft};\\
        &\frac{1}{\sqrt{\norm{t}}}=\int df\;\frac{1}{\sqrt{f}}\,e^{-i2\pi ft};\\
        &\frac{\sgn(t)}{\sqrt{\norm{t}}}=\int df\;\frac{i\,\sgn(f)}{\sqrt{f}}\,e^{-i2\pi ft},
    \end{split}
\end{equation}
we find that $\bm{q}(t)$ and $\bm{p}(t)$ are Hilbert transform pairs.

\section{Photon packets}
Inspired by the usual wave packets of frequency domain, we define ``photon packet'' modes of the form
\begin{equation}
    \bm{A}=\int dt\;\phi^*(t)\;\bm{a}_{\tau=0,+}(t),
\end{equation}
where $\phi$ is a narrow function, peaked at $t=0$, verifying
\begin{equation}
    \int dt\norm{\phi(t)}^2=1,
\end{equation}
so that
\begin{equation}
    \commut{\bm{A}}{\bm{A}^\dagger}=1.
\end{equation}

From these modes, we construct the usual Fock states
\begin{equation}
    \ket{N}_{\bm{A}}=\frac{\left(\bm{A}^\dagger\right)^N}{\sqrt{N!}}\ket{\vac};
\end{equation}
and coherent states
\begin{equation}
    \ket{\alpha}_{\bm{A}}=\exp\left[\alpha\,\bm{A}^\dagger-\alpha^*\,\bm{A}\right]\ket{\vac}.
\end{equation}

Using the commutator
\begin{equation}
    \commut{\bm{a}_{\tau,+}(t)}{\bm{A}^\dagger}=\phi(t-\tau),
\end{equation}
we get
\begin{equation}
    \bm{a}_{\tau,+}(t)\ket{N}_{\bm{A}}=\sqrt{N}\;\phi(t-\tau)\ket{N-1}_{\bm{A}}\;\forall N\geq1,
\end{equation}
and
\begin{equation}
    \bm{a}_{\tau,+}(t)\ket{\alpha}_{\bm{A}}=\alpha\;\phi(t-\tau)\ket{\alpha}_{\bm{A}}.
\end{equation}

We then get
\begin{equation}
    _{\bm{A}}\bra{N}\bm{v}_{\tau,+}(t)\ket{N}_{\bm{A}}=0;
\end{equation}
\begin{equation}
    _{\bm{A}}\bra{\alpha}\bm{v}_{\tau,+}(t)\ket{\alpha}_{\bm{A}}=\frac{1}{4\pi}\;\sqrt{\frac{Zh}{2}}\;\Im\left[\alpha\,\chi^*(t-\tau)\right],
\end{equation}
with
\begin{equation}
    \chi(t)=\int dt'\;\phi(t-t')\;\frac{1+i\,\sgn(t')}{\norm{t'}^{3/2}}.
\end{equation}

We also have
\begin{equation}
    \begin{split}
        &_{\bm{A}}\bra{N}\bm{v}_{\tau,+}(t_1)\bm{v}_{\tau,+}(t_2)\ket{N}_{\bm{A}}-\bra{\vac}\bm{v}_{\tau,+}(t_1)\bm{v}_{\tau,+}(t_2)\ket{\vac}\\
        =\;&\frac{Zh\,N}{64\pi^2}\;\Re\left[\chi(t_1-\tau)\,\chi^*(t_2-\tau)\right];
    \end{split}
\end{equation}
\begin{equation}
    \begin{split}
        &_{\bm{A}}\bra{\alpha}\bm{v}_{\tau,+}(t_1)\bm{v}_{\tau,+}(t_2)\ket{\alpha}_{\bm{A}}-\bra{\vac}\bm{v}_{\tau,+}(t_1)\bm{v}_{\tau,+}(t_2)\ket{\vac}\\
        =\;&\frac{Zh}{32\pi^2}\;\Im\left[\alpha\,\chi^*(t_1-\tau)\right]\,\Im\left[\alpha\,\chi^*(t_2-\tau)\right].
    \end{split}
\end{equation}

Interestingly, we find that $_{\bm{A}}\bra{\alpha}\bm{v}_{\tau,+}^2(t)\ket{\alpha}_{\bm{A}}-_{\bm{A}}\bra{\alpha}\bm{v}_{\tau,+}(t)\ket{\alpha}_{\bm{A}}^2=\bra{\vac}\bm{v}_{\tau,+}^2(t)\ket{\vac}$, just like in the frequency domain.

\begin{figure}
    \centering
    \includegraphics[width=0.8\textwidth]{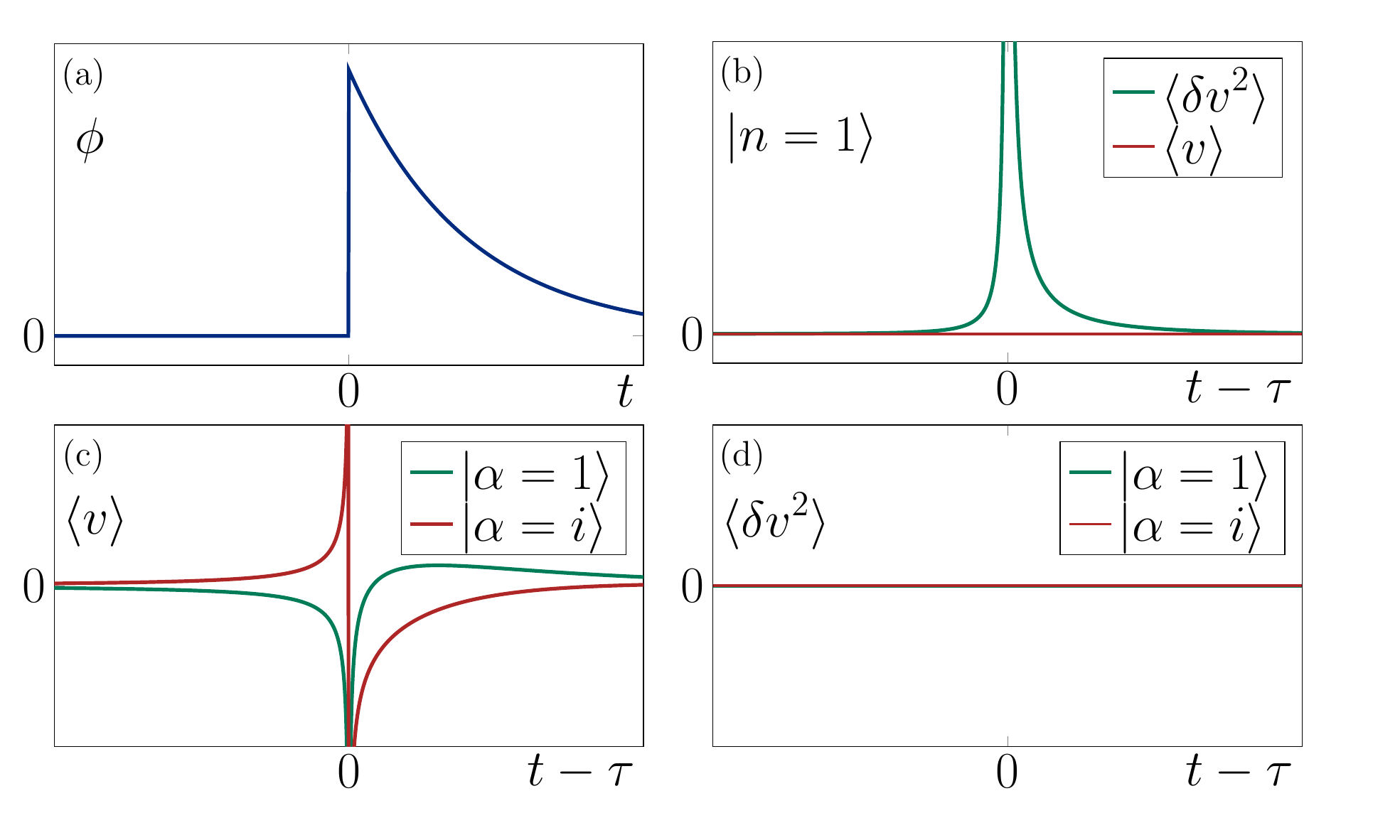}
    \caption{Traces of an exponential decay photon packet on an oscilloscope. (a): probability amplitude for the emission of a photon at location $\tau=0$, as a function of time; (b): resulting $\mom{v^2(t-\tau)}$ at location $\tau$ for the Fock state $\ket{n=1}$; $\mom{v(t)}$ is zero at all times for all Fock states; (c) and (d): resulting $\mom{v(t-\tau)}$ and $\mom{v^2(t-\tau)}$ at location $\tau$ for the coherent states $\ket{\alpha=1}$ and $\ket{\alpha=i}$. Time scales are identical on all figures.}
    \label{f6}
\end{figure}

The case of an exponentially decaying exponential photon packet amplitude is illustrated in Fig.~\ref{f6}.

The photonic amplitude probability of Fig.~\ref{f6}~(a) is
\begin{equation}
    \phi(t)=\frac{\Theta(t)}{\sqrt{\sigma_t}}\,e^{-t/2\sigma_t},
\end{equation}
where $\Theta$ is the Heaviside function, and we have
\begin{equation}
    \chi(t)=\frac{2i\sqrt{\pi}}{\sigma_t}\,e^{i\pi/4-\norm{t}/2\sigma_t}-2\sqrt{2}\,e^{-i\sgn(t)\pi/4}\left[\frac{1}{\sqrt{\norm{t}\sigma_t}}
    +\frac{\sqrt{2}\sgn(t)}{\sigma_t}\int_0^{\sqrt{\norm{t}/2\sigma_t}}du\;\exp\left(\norm{t}/2\sigma_t-u^2\right)\right],
\end{equation}
so that
\begin{equation}
    {\bm{A}}\bra{N}\bm{v}^2_{\tau,+}(t)\ket{N}_{\bm{A}}-\bra{\vac}\bm{v}^2_{\tau,+}(t)\ket{\vac}\simeq\frac{Z\hbar}{4\pi\norm{t-\tau}\sigma_t}
\end{equation}
around $t-\tau=0$.

\bibliographystyle{apsrev4-1}
\bibliography{UDTDQOv2_supp}